\documentclass[secnumarabic,amssymb,nobibnotes,aps,prd,notitlepage,longbibliography]{revtex4-1}

\usepackage{graphicx}
\usepackage{amsmath}

\graphicspath{{img/}}
\newcommand{\pdt}{\partial_t}
\newcommand\pd{\partial}
\renewcommand{\vec}[1]{\boldsymbol{\mathrm{#1}}} % Vector bold, as in JFM
\newcommand{\scr}[1]{\mathrm{#1}}

\usepackage{color}
\newcommand{\revAll}[1]{{\color{black}#1}}

\usepackage[normalem]{ulem}

\begin{document}

\title{Passive control of a falling sphere by elliptic-shaped appendages}

\begin{abstract}
The majority of investigations characterizing the motion of single or multiple particles in fluid flows consider canonical body shapes, such as spheres, cylinders, discs, etc. However, protrusions on bodies -- being either as surface imperfections or appendages that serve a function --  are ubiquitous in both nature and applications. In this work, we characterize how the dynamics of a sphere with an axis-symmetric wake is modified in the presence of thin three-dimensional elliptic-shaped protrusions.  By investigating a wide range of three-dimensional appendages with different aspect ratios and lengths,  we clearly show that the sphere with an appendage may robustly undergo an inverted-pendulum-like (IPL) instability. This means that the position of the appendage placed behind the sphere and aligned with the free-stream direction is unstable, in a similar way that an inverted pendulum is unstable under gravity. Due to this instability,  non-trivial forces are generated on the body,
%making it, for example, turn and drift,
leading to turn and drift,
if the body is free to fall under gravity. Moreover, we identify the aspect ratio and length of the appendage that induces the largest side force on the sphere, and therefore also the largest drift for a freely falling body. Finally, we explain the physical mechanisms behind these observations in the context of the IPL instability, i.e., the balance between surface area of the appendage exposed to reversed flow in the wake and the surface area of the appendage exposed to fast free-stream flow.

%We investigate the inverted-pendulum like (IPL) instability\cite{Lacis2014driftSymmetry} of sphere complemented with planar appendage shaped like an ellipse. We look into a range of aspect ratios, which characterize the shape of the appendage. For each appendage, we carry out a series of numerical simulations of a stationary body and evaluate the drag, lift and torque coefficients. We then compare the results for different appendages and conclude that there exists a shape, which yields an optimal performance with respect to lift force.  With this work, we want to initiate investigations, which would explore ways, in which a three-dimensional appendages can interact with surrounding flow in a beneficial way.
\end{abstract}

\author{U\v{g}is L\={a}cis$^1$, Stefano Olivieri$^2$, Andrea Mazzino$^{2,3}$ and Shervin Bagheri$^1$}%
\affiliation{$^1$Linn\'{e} Flow Centre, KTH Mechanics, SE-100 44 Stockholm, Sweden,\\ 
$^2$DICCA, University of Genova, 16145 Genova, Italy,\\
$^3$INFN and CINFAI Consortium, Genova Section, 16146 Genova, Italy}
%\email[Corresponding author: ]{shervin@mech.kth.se}%
\date{\today}%
\maketitle

\section{Introduction}
Organisms make use of  sophisticated passive control techniques by exploiting fluid-structure interaction instabilities and mechanisms \cite{fish2006passive}. One may divide these techniques into two categories; first, in which the fluid interacts with a complex surface (scales, hairs) distributed over a portion of a body, and second, in which the fluid interacts locally with isolated appendages (tails, pop-up feathers, etc). It is becoming clear that organisms can via both surfaces and appendages aid their locomotion (reducing drag/increasing lift), increase insulation (air-retention properties, heat transfer) or induce self-cleaning properties. How such properties are induced by the passive interaction of fluids and structures has only recently become known and is rapidly providing fuel for innovations. 
One example is represented by leading-edge tubercles inspired from the humpback whale \cite{miklosovic2004leading}. The wave-like modulation of the leading edge of an aerofoil is able to improve lift and drag characteristics as well as delay stall. Recently, tubercles has been also investigated as potential improvements for compressor aerofoils \cite{keerthi2015effect} and hydrofoils \cite{wei2015experimental}.

In this work, we focus on how a single three-dimensional appendage interacts with  steady and separated wake flows. We envision a passive control technique based on the precise design of body appendages in order to modify the force distribution around the appendage-less body  in a desired way. This paper presents a significant step towards this aim by  characterizing  how  thin elliptic-shaped appendages modify the pressure distribution around a sphere at a Reynolds number around 200. We focus, in particular, on a symmetry-breaking instability \cite{Lacis2014driftSymmetry} of the straight position (i.e., aligned with the incoming flow direction) of the appendage, which creates a significant side force, while often keeping the drag force essentially unmodified or even reduced.

This work illustrates that protrusions may have a significant effect on the path taken by free-falling objects by inducing additional instabilities
arising from non-trivial interactions with surrounding fluid.
%related to the protrusions.
As the review by Ern et al.\cite{ern2012wake} demonstrates, already canonical bodies (spheres, cylinders, disks, plates or bubbles) may have complex
%, non-trivial
falling/rising paths (tumbling, drifting, oscillating) depending on the density ratio.
The paths taken by canonical freely falling bodies are still actively researched~\cite{cano2016paths,cano2013wake,cano2016global}.
However, both in practical applications as well as in nature, bodies are not perfectly spherical, cylindrical etc; therefore the sensitivity of the falling paths to protrusions and corrugations of various shapes and sizes is an important -- but scarcely investigated -- issue.  

Protrusions are not only interesting to investigate because of imperfections, but also because they may serve a function of their own. The understanding of path generation mechanisms behind freely falling bodies can, for example, give important insights into seed dispersal. The focus on this work is on a protrusion-induced instability \cite{Bagheri_PRL_2012}, which has a significant effect on both falling and fixed bodies as explained by L\={a}cis et al.\cite{Lacis2014driftSymmetry}. In that paper it was shown that, when a sufficiently short splitter plate is attached to hind end of a free falling two-dimensional cylinder, the body turns and drifts. The instability of the straight position of the splitter plate was explained in \cite{Lacis2014driftSymmetry} by a semi-empirical model based on  an analogy to the instability of the upright position of an inverted pendulum system.
The same model has been used to understand the behaviour of very thin elastic filament
in a wake behind a three-dimensional circular cylinder by
Brosse et al.\cite{brosse2015experimental}.
The present work extends \cite{Lacis2014driftSymmetry} by characterizing three-dimensional elliptic-shaped appendages for a wide range of aspect ratios behind a sphere. The numerical results clearly show the presence of the inverted-pendulum-like (IPL) instability. Therefore, we will use the IPL model  to qualitatively explain the physical mechanisms behind the numerical results that are presented. In particular,  the IPL model provides the necessary physical intuition to explain - among other things -- why an appendage of a given shape and size provides a larger drift compared to other appendages.

This paper is organized as follows. In section \ref{sec:ipl-2d}, we  summarize the assumptions leading to the IPL model as well as the physical understanding gained by the model. This lays the foundation for explaining the  numerical results presented in remaining part of the paper. In section \ref{sec:opt-drift}, the parameters determining the geometry of the appendages are presented and the 
numerical flow solver is briefly explained as well as validated with respect to other works for the steady and axisymmetric flow behind a sphere at $Re=200$. 
Then, in section \ref{sec:ipl-all-angles}, we show the appearance of the IPL instability for two very different appendages; first (second) that mainly extends in the parallel  (transverse) direction of the free-stream. The behaviour of the forces on the bodies is explained and the role of the appendage shape with respect to the shape of the back-flow region of the wake is discussed. In section \ref{sec:ipl-all-lengths}, we conduct a parametric study over a range of appendage shapes and identify shapes that do not undergo IPL, shapes that undergo IPL instability and increase drag force and shapes that undergo IPL instability and essentially keep the drag unmodified (or even slightly reduced). The article is finalized with conclusions in section~\ref{sec:discuss}.

\section{Inverted-pendulum-like instability in two dimensions} \label{sec:ipl-2d}
\begin{figure}
  \begin{center}
  \includegraphics[width=0.8\linewidth]{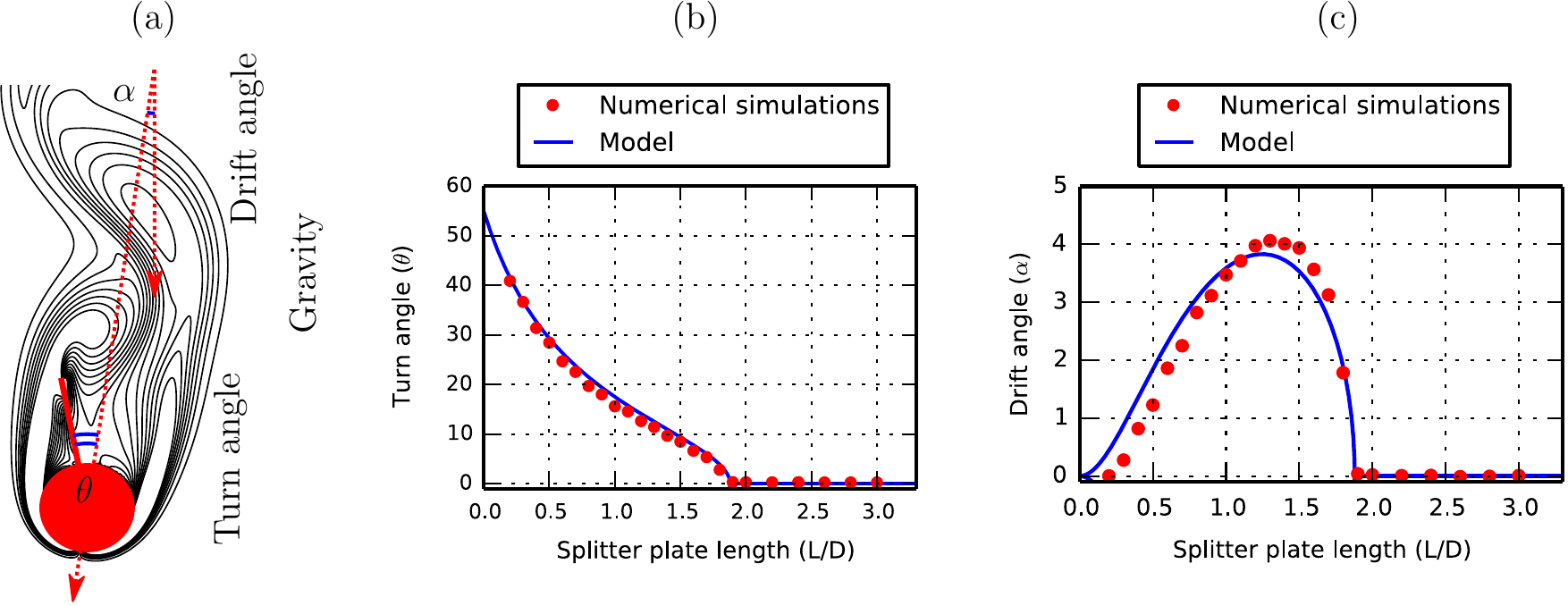} 
  \end{center}
  \caption{
Left frame (a) shows an instantaneous  snapshot of the flow around freely falling cylinder with splitter plate using  contour lines of the vorticity field.  Parameters are $\rho_\scr{s} / \rho_\scr{f} = 1.01$, $Re = 156$ and $L =D$. The body turns  $\theta = 19^{\circ}$ and drifts $\alpha = 8^{\circ}$ with respect to a vertical line. Frames (b) and (c) show, respectively,  turn and drift angles for different splitter plate  lengths for a freely falling body at $\rho_\scr{s} / \rho_\scr{f} = 1.001$ and $Re = 45$. The red dots correspond to numerical simulations, and the blue line to angles predicted by the IPL model. Figure adapted from
\cite{Lacis2014driftSymmetry}. \label{fig:def-2d-ipl-res}}
\end{figure}

We start by providing an example  taken from L\={a}cis et al.\cite{Lacis2014driftSymmetry} that demonstrates the inverted-pendulum-like (IPL) fluid-structure-interaction instability.  They considered a two-dimensional cylinder with a clamped rigid splitter plate  as illustrated in Fig.~\ref{fig:def-2d-ipl-res}$a$. The length of the splitter plate is $L = D$, where $D$ is the diameter of the cylinder.   The body density ratio with respect to the surrounding fluid is $\rho_\scr{s} / \rho_\scr{f} = 1.01$, where $\rho_\scr{s}$ is the density of the body and $\rho_\scr{f}$ is the density of the fluid.  When the body is released in a fluid at rest, it will, after a transient motion, reach a steady falling velocity $U_\scr{f}$. The Reynolds number based on $U_\scr{f}$ for this particular example is $Re = \rho_\scr{f} D U_\scr{f} / \mu = 156$, which means that the wake behind the body is unsteady.
A snapshot of vorticity isocontours around the falling body in the non-transient region is shown in Fig.~\ref{fig:def-2d-ipl-res}$a$, where the vortex shedding is clearly visible. Due to the presence of the splitter plate, the body is drifting to the left with an angle $\alpha = 8^{\circ}$. The body has also turned by an angle of $\theta = 19^{\circ}$ with respect to the direction of motion. Note that
 the trajectory is still oscillatory due to the von K\'{a}rman vortex street. The drift direction is always in the same direction as the splitter-plate is tilted, whereas the turn direction depends on the initial condition.  Figs.~\ref{fig:def-2d-ipl-res}$b$ and $c$  show how the drift and turn angles depend on the  length of splitter plate for $\rho_\scr{s} / \rho_\scr{f} = 1.001$ and $Re = 45$.  From both these plots one can observe that if the plate becomes shorter than the critical length $L_\scr{c} = 1.9\,D$, the body turns and a non-zero drift is generated. We also observe that for very short appendages, the turn angle approaches some finite value, which is close to the wake attachment angle $\theta_\scr{a} \approx 55^{\circ}$ for the cylinder alone. The drift angle, on the other hand, has a maximum value for intermediate plate lengths, and approaches zero drift for very small appendages. This is expected, since the cylinder alone does not exhibit any transverse motion.

\begin{figure}
  \begin{center}
  \includegraphics[width=0.6\linewidth]{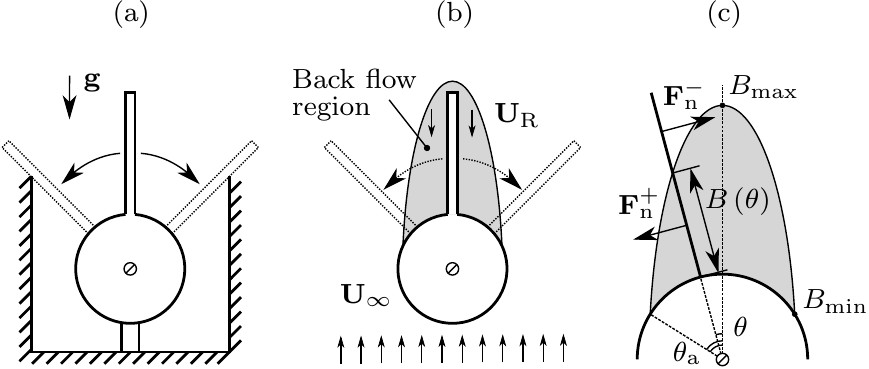} 
  \end{center}
  \caption{
Left frame (a) shows a sketch of an inverted-pendulum, where the vertical position is unstable. Frame (b) shows an analogous fluid problem, consisting of a cylinder and splitter plate exposed to incoming flow $\vec{U}_\infty$. Instead of the gravitational force, pressure force arising due to the reverse flow $\vec{U}_\scr{R}$ in the wake of the cylinder act as destabilizing force, turning the plate out of the back flow region, either to the right or to the left. In frame (c), the quantities defined within the IPL model are shown. The model of back flow region (MBFR) is shown in gray color. Figure adapted from
\cite{Lacis2014driftSymmetry}. 
 \label{fig:def-2d-ipl-mod}}
\end{figure}

\subsection{Analogy to an inverted pendulum}

The observations  above can be explained by a simple -- yet quantitative -- model based on an analogy to an  inverted pendulum confined between two vertical walls. The sketch in Fig.~\ref{fig:def-2d-ipl-mod}$a$ shows  a pendulum consisting of a circular cylinder and a plate. The body is free to rotate around the center of the cylinder, which is some distance below the center of mass of the pendulum. The straight vertical position of the pendulum is thus an unstable equilibrium  and any small disturbance will make it fall either to left or to the right. The   fluid-structure-interaction mechanism of the freely falling body is  similar; the only difference is that instead of gravitational forces the pressure forces are acting on the plate to destabilize it -- as illustrated in Fig.~\ref{fig:def-2d-ipl-mod}$b$.  In other words, if the splitter plate is sufficiently short, in the presence of a small disturbance the pressure forces will turn the splitter plate out of the back flow region (i.e., the region with a significant reversed flow behind the cylinder). However, as the plate is pushed out of the back flow region, it is exposed to forward flow, which provides a stabilizing force and acts in a similar way as the wall acts for the inverted pendulum (Fig.~\ref{fig:def-2d-ipl-mod}$a$).

\begin{figure}
  \begin{center}
  \includegraphics[width=0.4\linewidth]{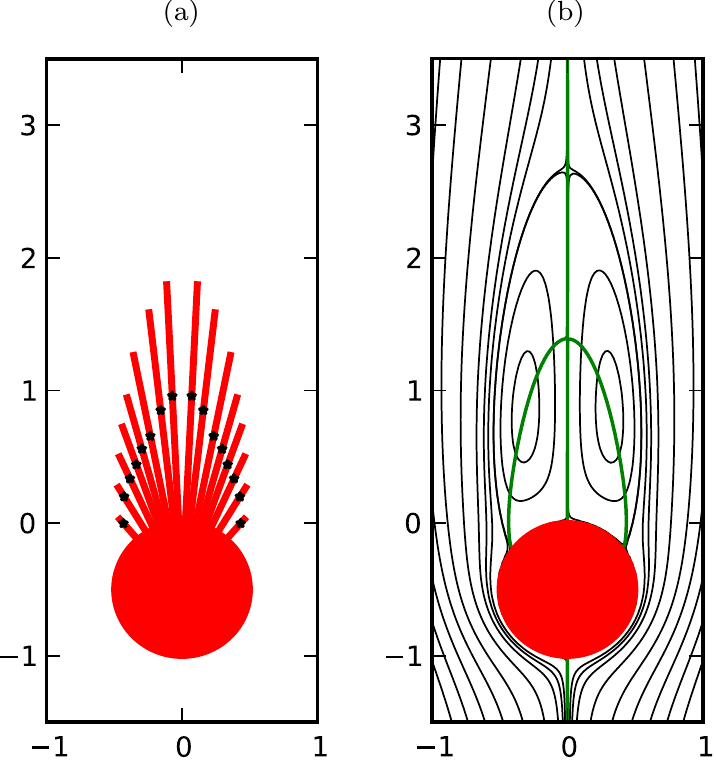} 
  \end{center}
  \caption{
Frame (a) shows results obtained from direct numerical simulations for fixed cylinder with splitter plate at equilibrium (zero-torque) turn angles. The point at which the normal force on the plate changes direction is marked using black star. Frame (b) shows the flow field obtained from numerical simulation around cylinder alone. Black lines are streamlines and the green line is contour line of zero azimuthal velocity $u_{\theta} = 0$. Figure adapted from
\cite{Lacis2014driftSymmetry}.\label{fig:def-2d-ipl-bfr-mod}}
\end{figure}

In order to obtain more quantitative predictions, a model of the back flow region can be defined as shown in Fig.~\ref{fig:def-2d-ipl-mod}$c$. Here,  $B\left( \theta \right)$ is the distance from the cylinder surface to the point on the plate where the normal force on the plate changes sign. Fig.~\ref{fig:def-2d-ipl-bfr-mod}$a$ shows these points (with black dots) on the plate identified from a series of simulations of the flow around cylinder with splitter plate at various equilibrium turn angles. It is however not always possible -- for example, in many experiments -- to directly evaluate the force distribution on the appendage. An estimate of the back flow region can be made from simulations or experimental measurements of the wake of a body {\it without} the appendage. This is based on the assumption that the thin appendage modifies the wake only locally. Under further assumptions provided in \cite{Lacis2014driftSymmetry},  the direction of the normal force is determined by velocity component normal to the plate, which in this set-up is the azimuthal velocity $u_{\theta}$.  Contour lines of zero azimuthal velocity $u_{\theta} = 0$ is shown with a green line in Fig.~\ref{fig:def-2d-ipl-bfr-mod}$b$.  One can see that the estimate obtained from  $u_{\theta} = 0$ condition provides the shape of the  back flow region, but overestimates its length. This condition, however, provides a more accurate estimate compared to recirculation bubble.
The length of the recirculation bubble can be obtained using negative stream-wise velocity at $x = 0$, which leads to a model back flow region approximately twice as big compared to the one predicted by the azimuthal velocity. 

Given a  model of back flow region (MBFR), the splitter plate is divided into two parts; the part inside the MBFR, where the normal force $F_\scr{n}^+$ acts in destabilizing direction or generates turn of the body; and the part outside of the MBFR, where the normal force $F_\scr{n}^-$ acts in stabilizing direction or opposes the turn of the body. The stabilizing force is defined as
\begin{equation}
F_\scr{n}^- = - 2 \sin(\theta)  C_A \rho_\scr{f} U_\infty^2  \left[ L - B\left(\theta\right) \right], \label{eq:2d-force-model-minus}
\end{equation}
where $\left[ L - B\left(\theta\right) \right]$ is the length of the plate outside of the MBFR. The parameter $C_A$ is the force law calibration coefficient. The destabilizing force  is defined as
\begin{equation}
F_\scr{n}^{+} = 2 k \sin(\theta)  C_A \rho_\scr{f} U_\infty^2\,  B\left(\theta\right),
\label{eq:2d-force-model-plus}
\end{equation}
where $B\left(\theta\right)$ is the length of the plate inside the MBFR. Here,  $k>0$ is a calibration coefficient describing the averaged magnitude of the force on the inner part of the splitter plate relative to the outer part. It can be shown that these force expressions are a special case of a commonly used model for describing the forces on a freely falling plate \cite{Andersen_JFM_2005_fallingPaperModel,Huang_PRevE_2013_ExperimetalFalling}.
Both calibration coefficients $C_A$ and $k$ can be estimated with reasonable accuracy from measurements of the wake without a splitter plate \cite{Lacis2014driftSymmetry}.

\subsection{Drift and turn angles}
Using the model of normal forces, we can obtain equilibria angles by finding zero torque around the center of the cylinder. The total torque  due to the splitter plate around the center of cylinder is 
\begin{equation}
T\left( \theta \right) = F_\scr{n}^{+} \left[\frac{D}{2} + \frac{B\left(\theta\right)}{2} \right] + F_\scr{n}^{-} \left[\frac{D}{2} + \frac{B\left(\theta\right)}{2}  + \frac{L}{2} \right] . \label{eq:torque0}
\end{equation}
To construct the torque expression, it has been assumed that the force on the plate is located at the center of each segment of the plate, as illustrated in Fig.~\ref{fig:def-2d-ipl-mod}$c$. Inserting the expressions for normal forces into (\ref{eq:torque0}), we obtain
\begin{equation}
T\left( \theta \right) =  \sin\left(\theta\right) \left\{ \left(1 + k \right) \left[ \hat{B}^2\left(\theta\right) + \hat{B}\left(\theta\right) \right] - \hat{L}^2 - \hat{L} \right\}  C_A \rho_\scr{f} U_\infty^2 D^2, \label{eq:torque}
\end{equation}
where 
$\hat{B} = B/D$ and $\hat{L} = L/D$. 
By choosing an appropriate value for $k$ based on appendage-free wake measurement, one can  find  equilibrium torque angles $\theta_0$ (i.e.,~angles for which $T(\theta_0)=0$) for each splitter plate length. The  results from the IPL model are compared to the DNS results of freely falling cylinder with splitter plate at $Re = 45$ in Fig.~\ref{fig:def-2d-ipl-res}$b$, where we observe that the agreement is  good. The  straight vertical position ($\theta_0 = 0^{\circ}$) looses it stability as the splitter plate length becomes shorter than a critical value $L_\scr{c}$. The {\it turn angle} $\theta_0$ thus refers to a non-trivial equilibrium angle of equation (\ref{eq:torque}) that exists for $L<L_\scr{c}$.

The forces on the splitter plate can be used to predict the total drift force acting on the whole body, which can be expressed as
\begin{equation*}
F_\scr{drift}\left(\theta_0\right) =  - \tilde{C}_A \cos \left ( \theta_0\right) \left(F_\scr{n}^+ + F_\scr{n}^- \right),
\end{equation*}
where $\tilde{C}_A$ is modified force calibration coefficient explained in \cite{Lacis2014driftSymmetry}. The associated {\it drift angle} \cite{vogel1994life} $\alpha$ is obtained from the ratio of the drift force to the drag force, i.e.
\begin{equation}
\alpha \simeq \arctan \left( \frac{F_\scr{drift}}{F_\scr{drag}} \right).
\end{equation}
%
%where drift $C_\scr{drift}$ and drag $C_\scr{drag}$ coefficients are defined as
%explained in paper \cite{Lacis2014driftSymmetry}.
The drift angle is positive $\alpha > 0$ if the turn angle is positive $\theta_0 > 0$ and vice-versa and in both cases the drift is in the direction to which the plate is turned. The drift angles obtained from the IPL model are  compared to DNS  of freely falling cylinder with splitter plate at $Re = 45$ in Fig.~\ref{fig:def-2d-ipl-res}$c$, where again a good agreement is observed. \revAll{The model plots in Fig.~\ref{fig:def-2d-ipl-res} has been produced using coefficient
values $\tilde{C}_A = F_\scr{drag}/(2 \rho_\scr{f} U^2_\infty D)$ and $k = 0.90$. The back flow region
is modelled as an half-ellipse attached to the cylinder
at angle $\theta_a = 55^{\circ}$ with the distance between the cylinder and the
tip of the ellipse $B\left(0\right) = 1.26\,D$. These coefficients, as
described in \cite{Lacis2014driftSymmetry}, has been calibrated with respect
to direct numerical simulations. The drag force $F_\scr{drag}$ is not explicitly
used, because it is cancelled out, when taking ratio between the
drift and drag forces.} In the remaining part of this work, we
explore the additional freedoms which a third dimension introduces by
considering various shapes of the appendages.

\begin{figure}
  \begin{center}
  \includegraphics[width=0.7\linewidth]{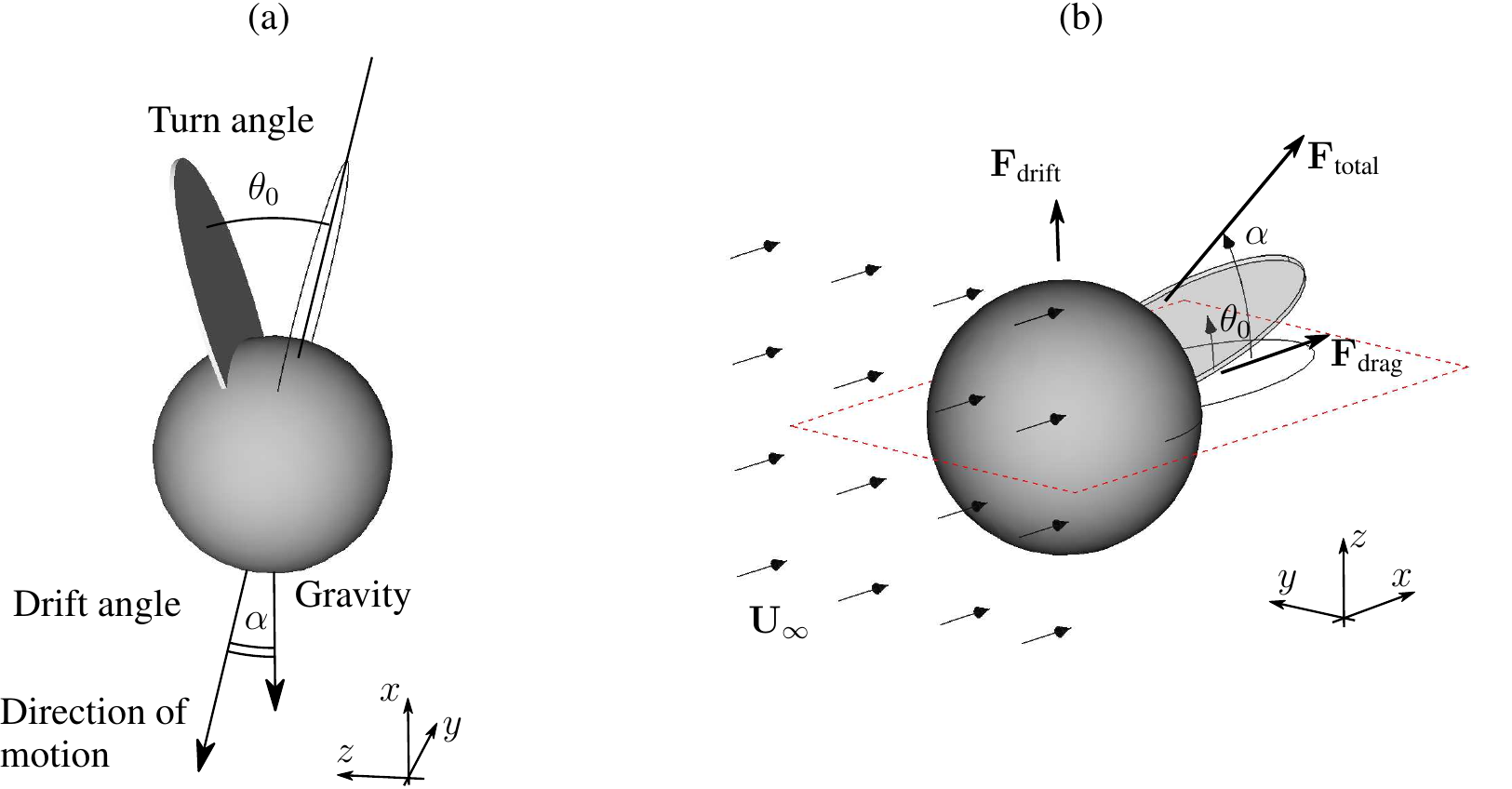} 
  \end{center}
  \caption{
Frame (a) shows the body consisting of a sphere and a planar, elliptical appendage (with aspect ratio $A=0.7$ and length $L=0.7\,D$) falling freely under the influence of  gravity.  Frame (b) shows the same body in a fixed framework, where the body is exposed to a constant free-stream velocity.  Figure (b) also defines the drift (or lift) force $\vec{F}_{\scr{drift}}$ and the drag force  on the body. The angle, $\alpha$,  between the total force and free stream direction in (b) corresponds to the drift angle in (a). The equilibrium angle $\theta_0$ in (b) corresponds to the turn angle in (a). \label{fig:def-3d-fall-prob}}
\end{figure}

\section{Three-dimensional configuration and numerical procedure} \label{sec:opt-drift}

\subsection{Geometry of a freely falling body}
Fig.~\ref{fig:def-3d-fall-prob}$a$ shows schematically the configuration of a sphere with an appendage  falling freely under gravity in still fluid. We are interested in characterizing the drift and turn angles, as defined in the figure, for appendages of different shapes and sizes. We define turn as rotation around the $y$-axis and drift as translation in the $z$ direction. 

\begin{figure}
\begin{center}
\includegraphics[width=1.0\linewidth]{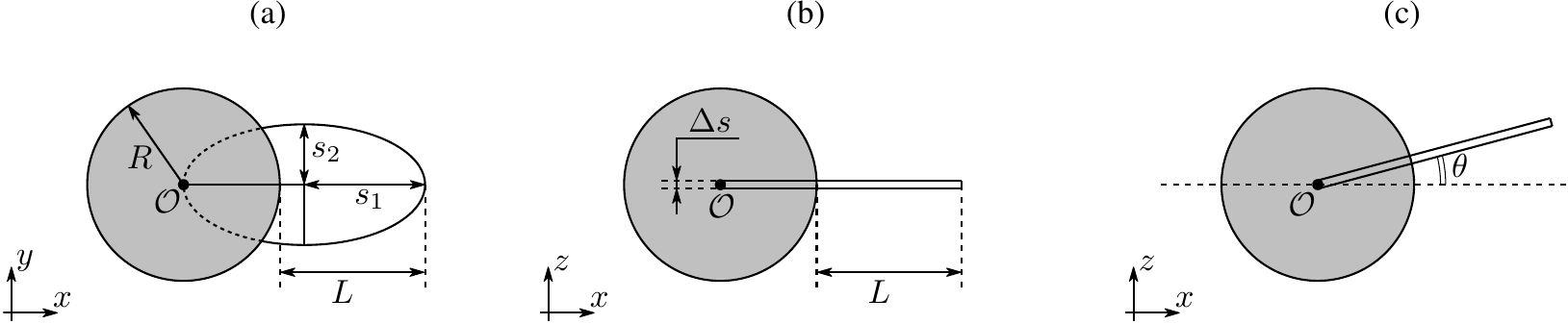} 
\end{center}
\caption{
Frames (a) and (b) show a top-view and side-view of the body under investigation, respectively.  The center of sphere and origin of the coordinate system is denoted by $\mathcal{O}$. Frame (c) shows the rotation of the body  around the $y$-axis by some turn angle $\theta$. \label{fig:def-object-w-app}}
\end{figure}

In order to take the first step in characterizing  appendage-induced instabilities, we make two major simplifications.  First, we limit ourselves to planar appendages that are shaped as ellipses with semi-major axis  $s_1$ and a semi-minor axis $s_2$ as shown in Fig.~\ref{fig:def-object-w-app}$a$. The parameters defining the geometry of the body  are thus the sphere diameter  $D=2R$, length of the appendage $L$ as measured from the back of the sphere, and the aspect ratio $A=s_2/s_1$ of the ellipse.  The elliptic appendage is aligned in such a way that one quadrant of the ellipse
 coincides with the center of the sphere. Therefore, the actual  appendage shape is only the part of the ellipse that extrudes the sphere.  Given these parameters, the length of ellipse semi-axis can be recovered as
\begin{equation*}
s_1 = \frac{1}{2} \left( \frac{D}{2} + L \right),
\end{equation*}
in the direction normal to the surface of the sphere and $s_2 = A\,s_1$ in the direction tangential to the surface of the sphere. The thickness of the appendage (Fig.~\ref{fig:def-object-w-app}$b$) will be kept constant $\Delta s = 0.02\,D$ with respect to the diameter of the sphere. 
 
The second simplification is related to the fact that the general problem as  illustrated in Fig.~\ref{fig:def-3d-fall-prob}$a$ is very challenging numerically, in particular at low density ratios \cite{lacis2016stable} where wake induced oscillations of the body exist \cite{ern2012wake}. In this work, we limit our investigations to sufficiently low Reynolds numbers such that the wake behind the sphere is steady but sufficiently large Reynolds numbers  such that a significant recirculation region exists, \revAll{i.e. $Re = 200$}. In this way, we avoid  non-trivial dynamic interactions between the motion of the body and the generated wake that may exist in time-dependent wakes. We can thus focus our attention on the instability generated by the appendage alone.  Therefore the  problem of freely falling body with a constant velocity is replaced with a fixed body exposed to a constant free-stream velocity as shown in Fig.~\ref{fig:def-3d-fall-prob}$b$. It thus follows that all the degrees of freedoms of the rigid-body dynamics are constrained; the rotation around $y$-axis is modelled by considering the body at various turn angles.
\revAll{
This approach is similar to the analysis carried out by
Fabre et al.\cite{fabre2011quasi,fabre2012steady}. They also consider steady
flow problems and relate the solution to the problem of freely falling body.
In order to investigate flow structures responsible for oblique falling paths,
they also carry out weakly non-linear expansion
in turn angle relative to the incoming flow velocity.
}

\subsection{Numerical procedure}\label{sec:setup}
The flow around a fixed body in an open domain exposed to uniform incoming free stream velocity $U_{\infty}$ is governed by the incompressible Navier-Stokes equations,
\begin{align*}
\rho_\scr{f} \left( \pdt \vec{u} + \left( \vec{u} \cdot \nabla \right) \vec{u}
\right) & =
 - \nabla p + \mu \Delta \vec{u}, \\
\nabla \cdot \vec{u} & = 0,
\end{align*}
where $\vec{u} = \left(u, v, w\right)$ is the  flow field and $p$ is the pressure field, both described in standard Eulerian coordinates. Here,
% $\rho_f$ is the fluid density and % It is defined before
$\mu$ is the fluid viscosity. The Reynolds number of the flow is defined based on  free-stream velocity and the sphere diameter, which is $Re = \rho_\scr{f} U_{\infty} D / \mu$. As mentioned before, we set the Reynolds number to $Re = 200$ throughout all this work.

% Here starts region, where Stefano made adjustments

To solve the equations above, we use the open-source  finite-volume based flow solver 
{OpenFOAM}~\cite{openfoamMain,Moukalled:2015:FVM:2876154}. 
The body is surrounded by a large rectangular box, with dimensions $x \in \left[-10D, 30D\right]$, $y \in \left[-10D, 10D\right]$ and $z \in \left[-10D, 10D\right]$. 
At the inlet plane $x = -10D$, a uniform free stream velocity is imposed as a Dirichlet boundary condition. At the outlet plane $x = 30D$, a convective outflow boundary condition is used, whereas at the lateral sides of the computational domain, slip boundary conditions are employed.
%According to our aim,
To obtain the solution,
we use the steady state solver \texttt{simpleFoam}. 
Central differencing scheme is employed for the diffusion term and gradients while the convective term is discretized using a second order upwind scheme.
The resulting linear systems arising from discretized equations are solved using
geometric-algebraic multi-grid (GAMG) and Gauss-Seidel solvers for pressure and velocity, respectively.
The pressure-velocity coupling is handled by an algorithm, which is known as the semi-implicit
method for pressure-linked equations (SIMPLE).
When the residual value becomes smaller than $10^{-6}$ (or after $2000$ iterations, by confirming
that the residual trends show convergence)
for both pressure and velocity, the iterative algorithm is stopped.
%Convergence is retained to be satisfactory when the residual value becomes smaller than $10^{-6}$, for both pressure and velocity, or after $2000$ iterations (monitoring that the residual trends are reasonable).

\begin{table}[b]
\begin{center}
\begin{tabular}{p{45mm} p{32mm} p{32mm} }
  & Drag coefficient $C_\scr{drag}$ & Length of wake $L_{\textrm{wake}}$  \\ \hline
Mesh A \revAll{($\Delta s = 0.025\,D$)} & $0.7371$ & $1.380$ \\
Mesh B \revAll{($\Delta s = 0.0125\,D$)} & $0.7463$ & $1.401$ \\
Mesh C \revAll{($\Delta s = 0.00625\,D$)}& $0.7546$ & $1.410$ \\
% Present results & $0.7463$ & $1.401$ \\
Johnson and Patel\cite{johnson1999flow} & $0.7759$ & $1.458$ \\
Gushchin and Matyushin\cite{gushchin2006vortex} & $0.7720$ & -- \\
Tomboulides et al.\cite{tomboulides1993direct} & -- & $1.431$ \\
\end{tabular}
\end{center}
\caption{Drag coefficient $C_\scr{drag}$ and wake length $L_{\textrm{wake}}$ are reported for the flow around fixed sphere at Reynolds number $Re = 200$ using three different meshes \revAll{(mesh spacing $\Delta s$ at the surface of the sphere is shown in the parenthesis)}. For comparison, numerical results by Johnson and Patel\cite{johnson1999flow},  Gushchin and Matyushin\cite{gushchin2006vortex} and 
Tomboulides et al.\cite{tomboulides1993direct} are also listed.
\label{tab:sphere-validation}}
\end{table}%

The computational domain is meshed with \texttt{cartesianMesh}, a mesh generator from the \texttt{cfMesh}~\cite{cfMesh} suite. It produces a predominantly hexahedral, body-fitted mesh with local refinement regions.
The grid resolution is refined closer to the body, using six refinement levels (from each level to the next one, the mesh spacing is halved). 
We consider three different cases with mesh spacing $0.8\,D$ (mesh A), $0.4\,D$ (mesh B)
and $0.2\,D$ (mesh C) at the outer boundaries of computational
domain.
We validate the numerical scheme using the flow around a sphere without an appendage. The obtained drag
coefficient $C_\scr{drag}$ and length of the wake $L_{\textrm{wake}}$ using all three meshes
are compared to literature~\cite{johnson1999flow, gushchin2006vortex, tomboulides1993direct} 
in Tab.~\ref{tab:sphere-validation}.
%Considering the sphere without appendage, validation is performed by comparing the obtained drag coefficient $C_\scr{drag}$ and length of the wake $L_{\textrm{wake}}$ to previous works~\cite{johnson1999flow, gushchin2006vortex, tomboulides1993direct}  in Tab.~\ref{tab:sphere-validation}.
Fig.~\ref{fig:bfr-pressure-dist}$a$ shows iso-surfaces of stream-wise velocities $u = 0$ and $u = -0.1 U_{\infty}$ of the wake behind the sphere. The length of the wake $L_{\textrm{wake}}$ is obtained by finding the distance from the surface of the sphere to the
tip of the blue $u = 0$ iso-surface.
% Tab.~\ref{tab:sphere-validation} shows that the agreement is satisfactory.
\revAll{
From Tab.~\ref{tab:sphere-validation} we observe that
%our results approach the values
%reported in literature~\cite{johnson1999flow, gushchin2006vortex, tomboulides1993direct}
%as we refine the mesh resolution.
the results obtained using the finest mesh resolution
are around $3\%$ below the drag and
wake length values reported
in literature~\cite{johnson1999flow, gushchin2006vortex, tomboulides1993direct}.
However, the trend of our results, as the mesh is refined, is towards the data
from literature. The change between different meshes is very small -- drag
and wake length change
by roughly $1\%$ after each reduction of the mesh size by a factor of two.
The convergence is slow due to the employed low-order finite-volume method.
For more accurate
results high-order methods,
such as spectral-element method\cite{karniadakis2013spectral}, should
be employed. This is, however, not the main aim of the present work.
For our purposes, we consider mesh B to be a satisfactory
compromise between accuracy and computational demands.}
This mesh, which will be used throughout the current work, consists of around $1.6 \cdot 10^6$ cells in total. The mesh spacing at the sphere is $\Delta x = \Delta y = \Delta z = 0.0125\,D$, whereas at the outer boundaries of the domain it is $\Delta x = \Delta y = \Delta z = 0.4\,D$.
Now we consider the sphere with an appendage. In order to verify that the numerical simulations are consistent, we carry out drag, lift and torque computations for sphere with appendage
using meshes A, B and C.
%in a similar way as for the sphere alone, we have carried out drag, lift and torque computations for sphere with
The appendage has aspect ratio $A = 0.2$ and length $L = 0.50\,D$; the body is turned using various turn angles.
% and using half and double mesh spacing as the production mesh.
The obtained drag, drift and torque coefficient values changed less than $2\%$ for the turn angles considered in this work. Therefore we have concluded that
the accuracy of our numerical scheme is sufficient to capture changes induced by
an addition of an appendage. 

% Here ends the region, where Stefano made changes

\begin{figure}
  \begin{center}
  \includegraphics[width=0.9\linewidth]{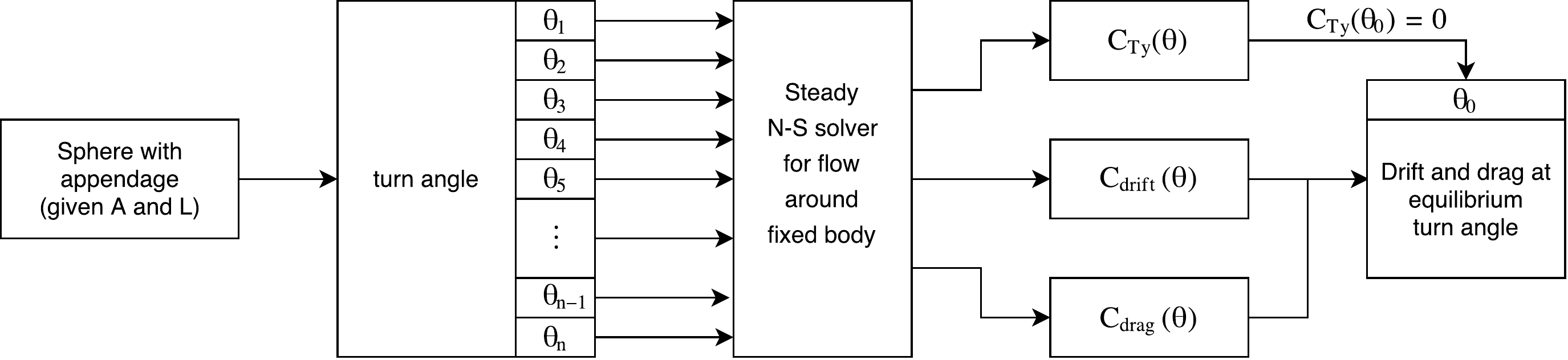} 
  \end{center}
  \caption{ This flow chart represents the simulation procedure for obtaining one point in the drift and turn angle plots. Starting with the given body (sphere with an appendage of fixed aspect ratio $A$ and length $L$), $n$ configurations with different turn angles around $y$ axis are constructed. Each configuration is investigated using steady Navier-Stokes solver. Combining the simulations, we obtain the torque, drift and drag coefficients as functions of the turn angle. The equilibrium angle $\theta_0$ is found when the torque coefficient is zero. \label{fig:sim-flow-chart}}
\end{figure}

To obtain the stable equilibrium angle arising due to IPL instability, a series of turn angles must be investigated in the fixed body framework, thus imitating the freedom of rotation around $y$ axis. The simulation procedure is schematically illustrated in Fig.~\ref{fig:sim-flow-chart} using a flow chart.
%If not mentioned otherwise, we
Given aspect ratio $A$ and appendage length $L$, we
%investigate turn angles $\theta \in \left[ -15^{\circ}, 15^{\circ} \right]$
carry out $n$ steady
simulations for turn angles $\theta \in \left[ -40^{\circ}, 40^{\circ} \right]$
using step of $\Delta \theta = 4^{\circ}$. 
%In this angle interval,
%and with this sampling distance,
%we carry out a number of steady simulations for the given aspect ratio $A$ and appendage length $L$.
From each simulation, we compute three integral observables, 
\begin{equation}
C_\scr{drag} = \frac{F_\scr{drag}}{\frac{1}{2}\rho_f U_{\infty}^{2} \frac{\pi}{4} D^2}, \quad
C_\scr{drift} = \frac{F_\scr{drift}}{\frac{1}{2}\rho_f U_{\infty}^{2} \frac{\pi}{4} D^2}
\qquad \mbox{and} \quad
C_{Ty} = \frac{T_y}{\frac{1}{2}\rho_f U_{\infty}^{2} \frac{\pi}{4} D^3},
\label{eq:def-drift-torque-coef}
\end{equation}
corresponding to the drag coefficient $C_\scr{drag}$, the drift (or lift) coefficient $C_\scr{drift}$ and the torque coefficient around $y$-axis, respectively. Specifically, the forces $F_\scr{drag}$ and $F_\scr{drift}$ are defined by
\begin{equation}
F_\scr{drag} = \left[ \int\limits_{\pd \Omega_s} \mathbf{\tau} \cdot \vec{n}\,\mathit{dS} \right ]
\cdot \hat{e}_x, \qquad \mbox{and} \qquad
F_\scr{drift} = \left[ \int\limits_{\pd \Omega_s} \mathbf{\tau} \cdot \vec{n}\,\mathit{dS} \right ]
\cdot \hat{e}_z ,
\end{equation}
where $\mathbf{\tau}$ is the fluid stress tensor and $\pd \Omega_s$ is the surface of the body. The torque is obtained from
\begin{equation}
\quad
T_{y} = \left [ \int\limits_{\pd \Omega_s} \vec{r} \times \left( \mathbf{\tau} \cdot \vec{n} \right)\,\mathit{dS} \right ] \cdot \hat{e}_y,
\end{equation}
where $\vec{r}$ is the radius vector, pointing from the center of the body to each point on the body surface. For simplicity, we have assumed that the appendage is so thin that the center of the mass for the sphere with an appendage coincides with the center of sphere itself. The torque is evaluated with respect to the axis that goes through the center of the sphere. 

\section{IPL instability and equilibrium solutions} \label{sec:ipl-all-angles}
\subsection{Forces and drift angle for an awl-like appendage ($A=0.7$, $L=0.7D$)} \label{sec:awl}
\begin{figure}
  \begin{center}
  \includegraphics[width=1.0\linewidth]{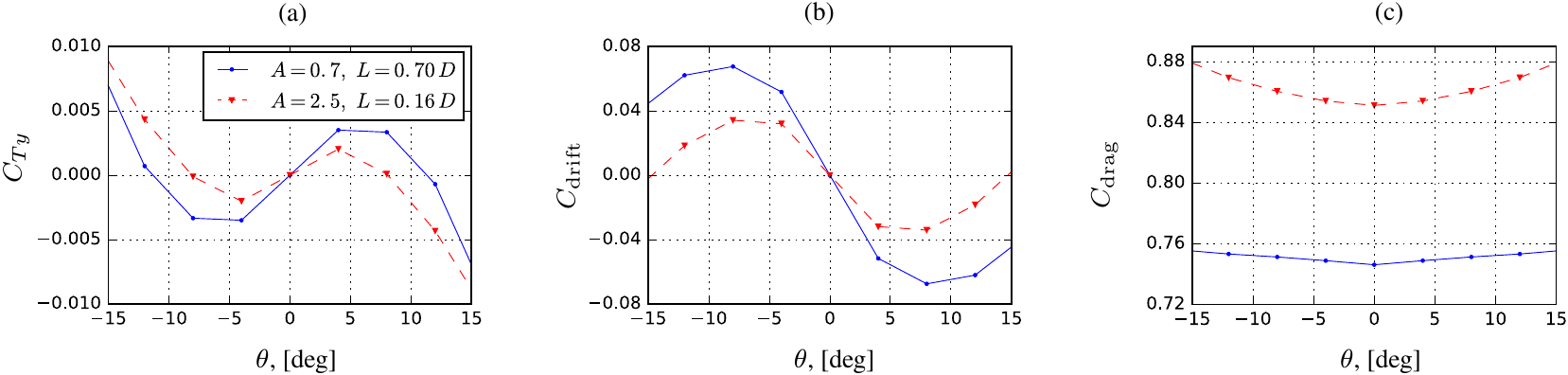} 
  \end{center}
  \caption{
Frame (a) shows the torque coefficient $C_{Ty}$ around the $y$-axis as a function of the turn angle. The blue and the red triangles correspond to the parameters $A = 0.7$, $L = 0.7\,D$ and $A = 2.5$, $L = 0.16\,D$, respectively. Frames (b) and (c) show the drift coefficient $C_\scr{drift}$ and drag coefficient $C_\scr{drag}$, respectively. Each marker in the plot corresponds to one steady numerical simulation. \label{fig:single-body-torqLiftDrag}}
\end{figure}

\begin{figure}
\begin{center}
  \includegraphics[width=1.0\linewidth]{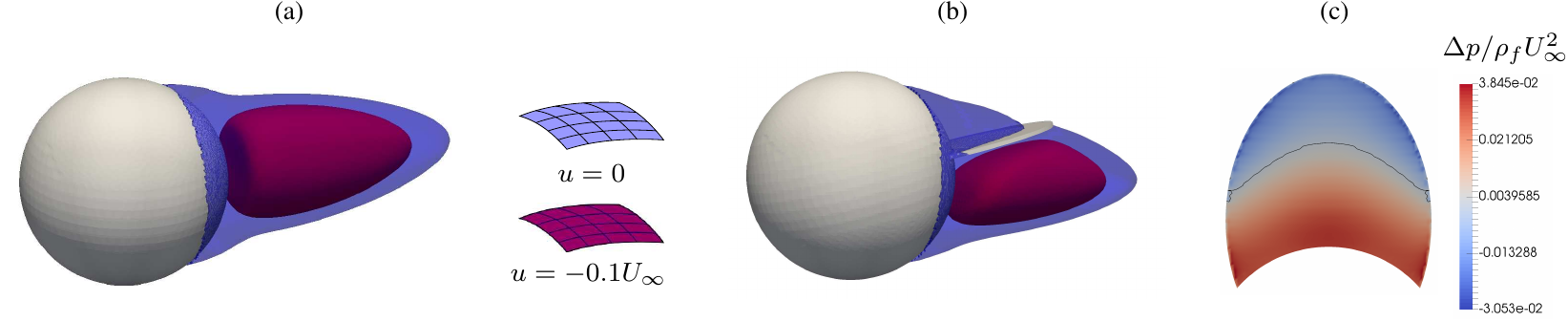} 
  \end{center}
  \caption{
Frames (a) and (b) show the wake behind the sphere at $Re=200$ without and with appendage ($L = 0.7\,D$, $A = 0.7$, $\theta = -11.3^{\circ}$), respectively. \revAll{Note that the flow field reported in the frame (b) has been obtained by solving additional steady problem at the determined equilibrium turn angle $\theta = -11.3^{\circ}$.} The blue iso-surface corresponds to zero streamwise velocity $u = 0$, while the red iso-surface corresponds streamwise velocity $u = -0.1 U_{\infty}$. The blue iso-surface has been made transparent to better observe the encompassed negative streamwise velocity iso-surface. Frame (c) shows the pressure difference -- between bottom and top of the appendage -- distribution over the appendage; zero contour-line ($\Delta p = 0$) is depicted by a black line.}
\label{fig:bfr-pressure-dist}
\end{figure}

We begin with illustrating the IPL instability on an appendage with aspect ratio $A = 0.7$ and length $L = 0.7\,D$, as shown in Figs.~\ref{fig:def-3d-fall-prob},
\ref{fig:bfr-pressure-dist}$b$ and 
\ref{fig:single-app-turnDriftVsLength}$a$ (middle). Using the numerical procedure described in the previous section and shown in Fig.~\ref{fig:sim-flow-chart}, we obtain the torque coefficient $C_{Ty}$, the drift coefficient $C_\scr{drift}$ and drag coefficient $C_\scr{drag}$ for different values of the turn angle; the results are shown in the interval $\theta \in \left[ -15^{\circ}, 15^{\circ} \right]$ with a blue line in Figs.~\ref{fig:single-body-torqLiftDrag}$a$,$b$~and~$c$, respectively.
The drag coefficient (Fig.~\ref{fig:single-body-torqLiftDrag}$c$) exhibits a minimal value at zero turn angle $\theta = 0^{\circ}$, i.e. $C_{\scr{drag}} = 0.7462$, which is very close to the value of the sphere alone.  We observe that the drag increases with the turning angle linearly. However, the increase is roughly 1\%, which means the drag force is rather insensitive to the appendage in this range of angles.  

Next, we provide a physical explanation of the torque and drift force on the body as the body is turned towards negative angles. If the body is turned towards positive angles, the same mechanism takes place but in the opposite direction. Therefore the plots in Figs.~\ref{fig:single-body-torqLiftDrag}$a$ and $b$ are antisymmetric. When the turn angle is zero $\theta_0 = 0^{\circ}$,  there is zero torque (equilibrium solution) and zero drift force, which is expected, since the flow behind the sphere with a straight appendage at $Re=200$ is symmetric with respect to plane $z = 0$. As the body is turned towards negative angles $\theta\rightarrow 0^-$, the pressure force below the appendage (i.e. the side of the appendage nearest $z=0$ plane) is larger than the pressure force acting from the upper side of the appendage. This results in a net normal force on the plate that would turn the plate further away from the straight $\theta=0$ position and towards negative angles, $\theta<0$, if rotation would be allowed. This is manifested by the negative torque around the $y$-axis for $\theta\rightarrow 0^-$  shown in Fig.~\ref{fig:single-body-torqLiftDrag}$a$.  The positive drift force (Fig.~\ref{fig:single-body-torqLiftDrag}$b$) is produced by the same mechanism after the force is projected onto the $z$-axis. Roughly speaking, the normal force on the plate is proportional to $\rho (\mathbf{U}_R\cdot\hat e_n)^2$, where $\hat e_n$ is normal vector of the plate surface. The net normal force arises from the fact that the reversed velocity projected in the direction of $\hat e_n$ is larger below the surface than above it. The relatively fast reversed velocity in the back flow region towards the bottom of the appendage can be seen in Fig.~\ref{fig:bfr-pressure-dist}$b$ when $\theta = -11.3^{\circ}$. 

As the appendage is turned towards negative $\theta$, the normal force from the upper side of the appendage grows, because part of the appendage (the tip) protrudes the region containing a reversed flow and is exposed to very fast forward flow (see again Fig.~\ref{fig:bfr-pressure-dist}$b$).
%We show
In Fig.~\ref{fig:bfr-pressure-dist}$c$ we show the pressure difference between the lower surface  (smaller $z$ values) and upper surface (larger $z$ values)  of the appendage. One can clearly see that the normal force changes  direction at the zero pressure difference iso-contour (black line). When the normal force above the plate has grown sufficiently large, the negative torque becomes positive and the plate is pushed back towards $z=0$ plane. The positive torque for $\theta < -11.3^{\circ}$ can be observed in Fig.~\ref{fig:single-body-torqLiftDrag}$a$. 
Note that although the torque changes sign, the drift and drag forces remain positive for all negative angles in the range shown in Fig.~\ref{fig:single-body-torqLiftDrag}$b$. The net normal force is therefore in the same direction, but the torque changes sign because the lever arms to positive and negative normal forces are different. In other words, the magnitude and surface area of the pressure difference distribution in the $+$ region is larger compared to the $-$ region, which results in a larger force on the $+$ region compared to the $-$ region, as illustrated schematically in
%Fig.~\ref{fig:bfr-pressure-dist}$c$.
Fig.~\ref{fig:illustrate-plusMinus-smallLargeA}$b$.

From the torque coefficient plot Fig.~\ref{fig:single-body-torqLiftDrag}$a$ one can observe that -- in addition to $\theta=0^{\circ}$ -- there are two  equilibrium (zero-torque) points, which are $\theta_0 = -11.3^{\circ}$ and $11.3^{\circ}$.
%\revAll{obtained using linear interpolation between neighbouring angles, which were computed.
%Note that to obtain the flow fields in Figs.~\ref{fig:bfr-pressure-dist}$b$ and
%\ref{fig:bfr-pressure-dist}$a$, we have solved additional steady problem at the obtained
%equilibrium turn angles.}
At the equilibria, the torque from these forces should be roughly balanced due to difference of the level arm. Having identified three equilibrium angles ($\theta=0^{\circ},\pm 11.3^{\circ}$) governed by zero-torque condition, one has to determine, if the states are stable solutions. The stability condition for the equilibrium turn angle is
\begin{equation*}
\left. \frac{\pd C_{Ty}}{\pd \theta} \right|_{\theta=\theta_0} < 0,
\end{equation*}
which essentially states that if the body is rotated away from the equilibrium angle by some external perturbation, a restoring torque will appear, which will turn the body back to the equilibrium angle. If this condition is not met, then the equilibrium is unstable, i.e., any perturbation on the turn angle will cause torque, which would turn the body even further away from the unstable equilibrium turn angle.
\revAll{
One has to recall the simplification, in which the freely falling
body is replaced with a
fixed body at different turn angles. Therefore
dynamic instabilities, which would
not be exposed by the current investigation technique, can not in principle
be excluded.
%However, authors believe that in the
%present parameter regime there are no dynamical instabilities.
}

For the current appendage torque (Fig.~\ref{fig:single-body-torqLiftDrag}$a$), equilibrium angles $\theta_0 = -11.3^{\circ}$ and $11.3^{\circ}$ fulfil the \revAll{static} stability condition, while the aligned equilibrium angle $\theta_0 = 0^{\circ}$ is unstable. After determining the equilibrium lift and the drag coefficients $C_\scr{drift0}$ and $C_\scr{drag0}$, we find the
corresponding drift angle $\alpha$ from
\begin{equation*}
\alpha = \arctan\left( \frac{F_\scr{drift0}}{F_\scr{drag0}} \right) = \arctan\left( \frac{C_\scr{drift0}}{C_\scr{drag0}} \right) = \arctan\left(
\frac{0.06304}{0.7529} \right) = 4.8^{\circ},
\end{equation*}
which is also the angle formed between the vertical direction and the path which the constructed body would follow if allowed to freely fall. Note that the drift angle does not depend of the drift and drag coefficient normalization factors introduced in expressions (\ref{eq:def-drift-torque-coef}), such as the projected area.

\subsection{Forces and drift angle for a wide appendage ($A=2.5$, $L=0.16\,D$)}
\begin{figure}[b]
  \begin{center}
  \includegraphics[width=1.0\linewidth]{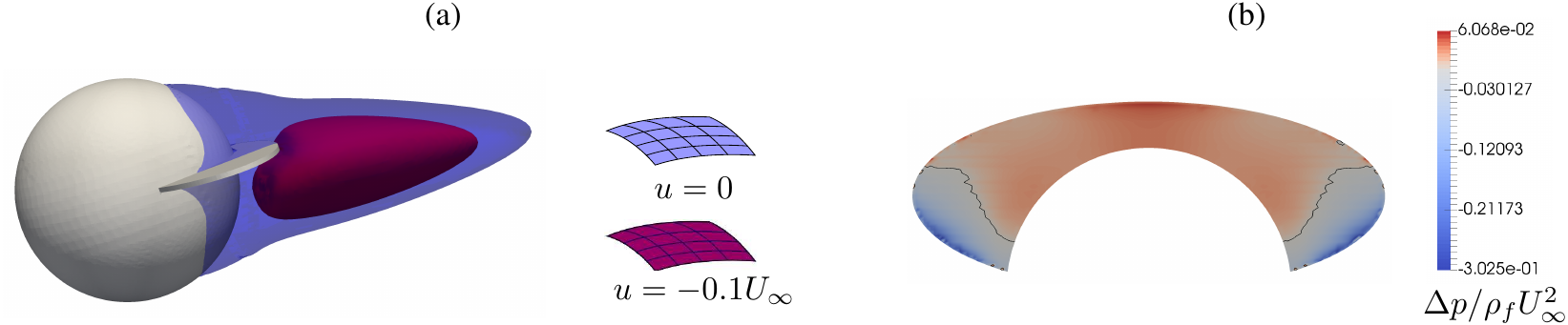} 
  \end{center}
  \caption{Frame (a) shows the wake behind the sphere with appendage ($A = 2.5$, $L = 0.16\,D$, $\theta = -8.1^{\circ}$) at $Re = 200$. \revAll{Note that the flow field reported in the frame (a) has been obtained by solving additional steady problem at the determined equilibrium turn angle $\theta = -8.1^{\circ}$.} Frame (b) shows the pressure difference distribution over this appendage with zero ($\Delta p = 0$) contour-line depicted by a black line.
\label{fig:bfr-pressure-dist3}}
\end{figure}

Next, we consider a wide and short appendage as shown in
Figs.~\ref{fig:bfr-pressure-dist3}$a$ and \ref{fig:all-app-turnDriftVsLength}$j$,
for which the aspect ratio is $A=2.5$ and length is $L=0.16\,D$. This appendage experiences the same instability as the awl-shaped appendage ($A=0.7$) discussed previously and can be physically understood in a similar way. However, because the appendage is wider than the sphere, the whole body experiences a significantly higher drag force than the sphere alone. The drag coefficient, shown in Fig.~\ref{fig:single-body-torqLiftDrag}$c$  (red color), increases more rapidly with the angle compared to awl-shaped appendage, because the surface area exposed to the free-stream increases.

The torque and drift force coefficients are qualitatively similar to the awl-shaped appendage, i.e. anti-symmetric with respect to $\theta=0$. The difference is that the wide appendage protrudes the back flow region from the lateral sides.  Fig.~\ref{fig:bfr-pressure-dist3}$b$ shows the difference between the pressure above and below the plate. We observe that the destabilizing pressure force covers nearly the entire appendage, except for two patches on the sides that are exposed to the fast free stream and therefore also to a stabilizing force. From Fig.~\ref{fig:single-body-torqLiftDrag}$b$, we clearly see that the drift force is positive for negative angles (as for $A=0.7$). For this appendage however, the direction of the drift force cannot be explained by arguing that the lever-arm to the positive pressure region is shorter than the lever-arm to the negative pressure region. The IPL moded developed for 2D bodies and summarized in section \ref{sec:ipl-2d} is based on the assumption that the modification of the back flow region due to the presence of the appendage is small, which is not the case for such a wide appendage as investigated here. In other words, the pressure distribution around the sphere is significantly modified due to the wide appendage, which means that, when determining the direction of the drift, it is not sufficient to characterize forces on the appendage alone and the side force on the sphere should be investigated.
 
For the wide appendage, we find two equilibria at turn angles $\theta=\pm 8.1^\circ$ that satisfy the stability condition. This body, if it were to fall freely under gravity, would therefore drift at an angle of $2.2^\circ$. We thus note that wide and short appendages do not exploit the IPL instability as efficiently as the slender awl-like appendage for generating drift. In the following section, we investigate a range of aspect ratios and appendage lengths in order to find the  appendage that induces maximum drift of the whole body.

\section{Appendages for IPL instability and largest drift } \label{sec:ipl-all-lengths}

\subsection{Critical and optimal lengths for awl-shaped appendage ($A=0.7$)}
\begin{figure}
  \begin{center}
  \includegraphics[width=1.0\linewidth]{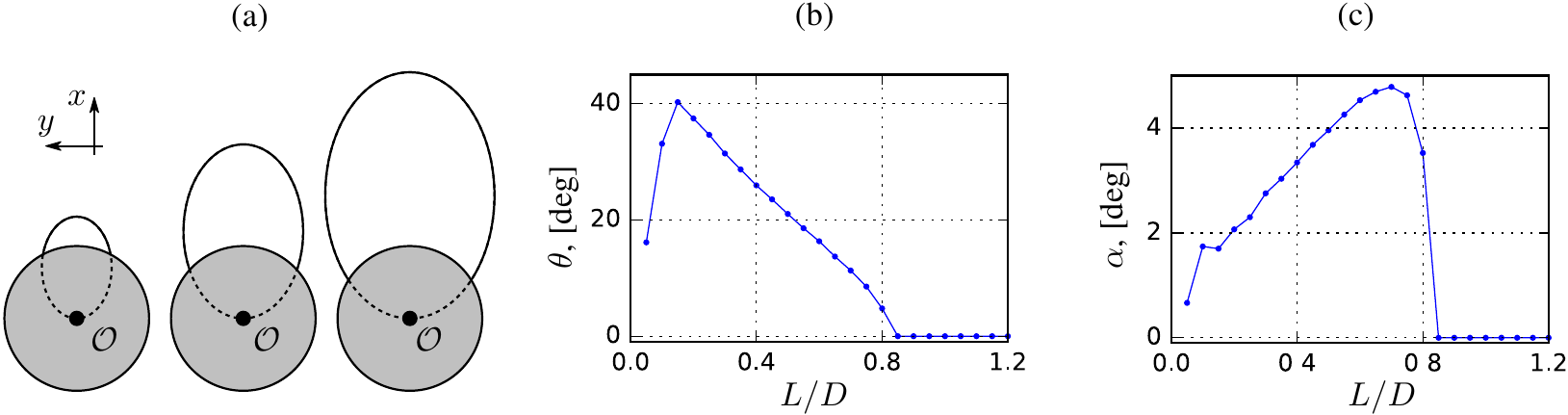} 
  \end{center}
  \caption{
Frame (a) shows the appendages behind sphere with aspect ratio $A = 0.7$ and lengths $L = 0.2\,D$, $0.7\,D$ and $1.2\,D$ (from left to right), sliced at $z = 0$ with plane $(x,y)$. Frames (b) and (c) show, respectively, the turn and drift angles for all appendage lengths investigated. Each marker in turn and drift angle plot corresponds to $n$ steady numerical simulations.
\label{fig:single-app-turnDriftVsLength}}
\end{figure}

Going back to the appendage with $A=0.7$ discussed in section \ref{sec:awl}, we now continue by characterizing the turn and drift angles for different lengths of the appendage. We start from $L = 0.05\,D$ and increase the length by steps of $\Delta L = 0.05\,D$ until we arrive with lengths, at which the IPL instability is not present any more.  Fig.~\ref{fig:single-app-turnDriftVsLength}$a$ shows the appendage  -- sliced at the $(x,y)$ plane at a zero turn angle -- with fixed aspect ratio $A = 0.7$ and lengths $L = 0.2\,D$, $0.7\,D$ and $1.2\,D$ (from left to right, respectively).
In Fig.~\ref{fig:single-app-turnDriftVsLength}$b$, we show the obtained turn angles for all of the appendage lengths considered for the aspect ratio $A = 0.7$. Note that for each $L$, the stable equilibrium turn angle is obtained by running a series of turn angle simulations -- as described in section~\ref{sec:ipl-all-angles} and flow chart in Fig.~\ref{fig:sim-flow-chart}. If the equilibrium turn angle is not found in interval $\theta \in \left[ -40^{\circ}, 40^{\circ} \right]$, the computational angle interval is extended to encompass the
stable equilibrium. From the Fig.~\ref{fig:single-app-turnDriftVsLength}$b$, we can conclude that similarly as for the 2D bodies
(see Fig.~\ref{fig:def-2d-ipl-res}$c$), if the appendage is longer than some critical length $L_c$, only the straight,
free-stream aligned position of the body is stable. For the current aspect ratio,  the critical length is
\begin{equation*}
L_c = \left( 0.825 \pm 0.025 \right)\,D.
\end{equation*}
However, if the appendage is shorter than $L_c$, there appears non-zero equilibrium turn angles via a pitchfork bifurcation. Note that the symmetric branch for the negative angles also exists but is not shown. As the appendage length is decreased further, the turn angle increases, in a similar fashion as for the plate behind circular cylinder (see Fig.~\ref{fig:def-2d-ipl-res}$c$). For the two shortest appendage lengths the turn angle is reduced, which is a behaviour that has not been observed for two-dimensional body. This reduction can be explained by the fact that the length of the appendage starts to become comparable with the thickness of the appendage ($\Delta s = 0.02\,D$), in which case it can not be viewed as a planar appendage in the back flow any more and it is likely that the efficiency of capturing the pressure forces in the wake is severely reduced.

\begin{figure}
  \begin{center}
  \includegraphics[width=1.0\linewidth]{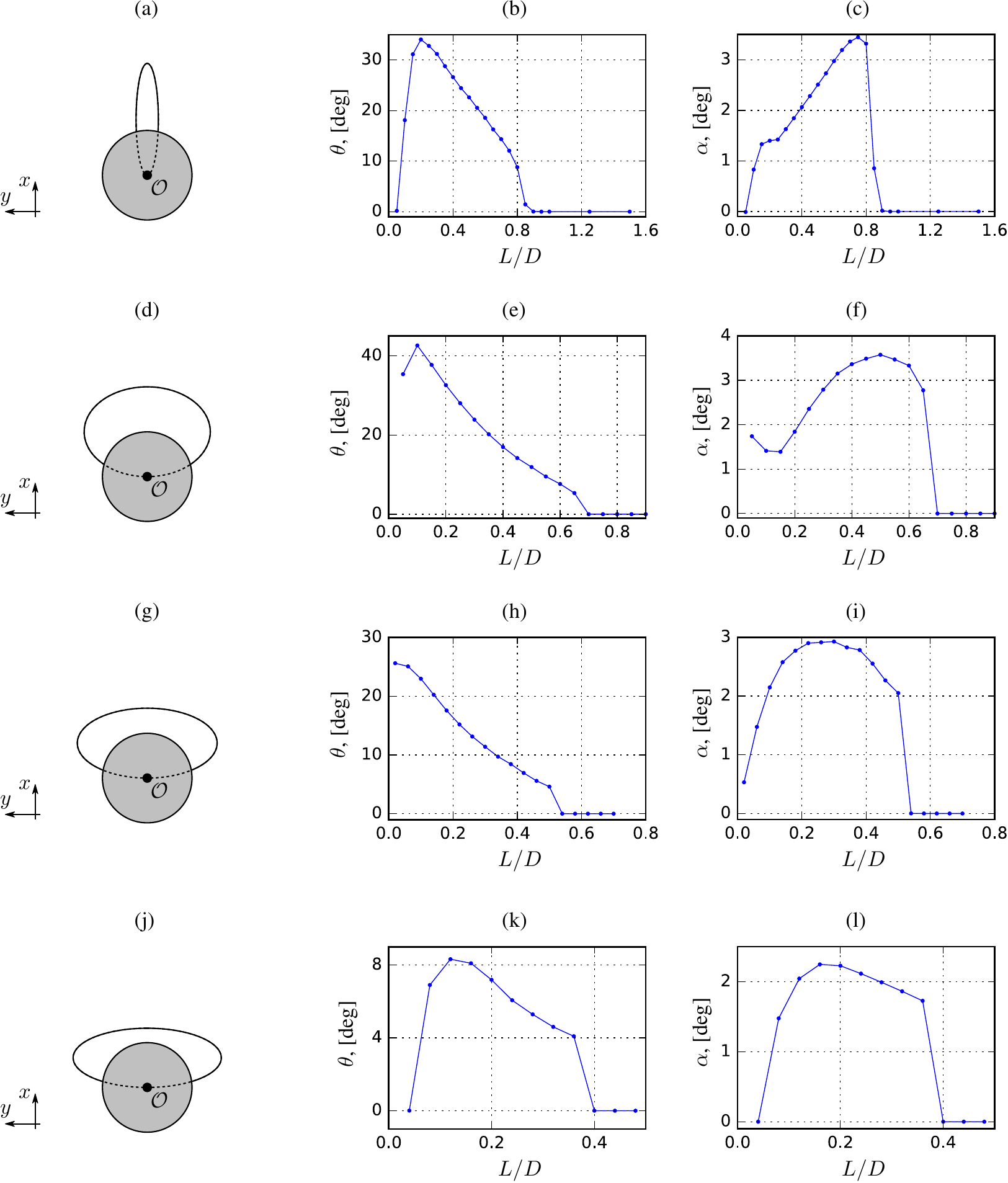} 
  \end{center}
  \caption{
Turn and drift angle curves for  different appendage shapes are shown. In frames (a,d,g,j) we show the appendage shape with aspect ratios $A = 0.2$, $1.4$, $2.0$ and $2.5$ and appendage lengths $L_{\max} = 0.75\,D$, $0.50\,D$, $0.28\,D$ and $0.16\,D$, respectively. The length $L_{\max}$ yields the largest drift. The second and third column show turn angle (b,e,h,k) and drift angle (c,f,i,l), respectively. 
\label{fig:all-app-turnDriftVsLength}}
\end{figure}

In Fig.~\ref{fig:single-app-turnDriftVsLength}$c$, we show the obtained drift angles for the considered appendage lengths. Similarly as for the turn angle, one observes that for appendage lengths $L > L_c$ the problem is trivial -- there is no drift angle present. However, for shorter appendage lengths $L < L_c$ the IPL instability appears and drift is generated. The drift angle quickly reaches an optimal value $\alpha = 4.8^{\circ}$ at $L_{\max} = 0.7\,D$, i.e., the largest drift for the current appendage shape. For shorter lengths, the drift angle is nearly monotonically decaying. Small values of $\alpha$ for the long appendage $(L\simeq 0.8 D)$ can be explained by its small turn angle, that is, there is a very small normal force generated on the appendage at small $\theta$ according to two-dimensional force model (Eqs.~\ref{eq:2d-force-model-minus}--\ref{eq:2d-force-model-plus}). Small values of $\alpha$ for shorter appendages ($L\lesssim 0.4 D$), on the other hand, can be explained by the small surface area, with which the appendage can interact with the surrounding back and forward flow. The small ``kink'' at $L = 0.10\,D$ can be attributed to the reduction in the turn angle, i.e., slightly reducing turn angle going from $L = 0.15\,D$ to $L = 0.10\,D$ leads to better efficiency in generating the drift force.

\subsection{Critical and optimal lengths for appendages $A= \{0.2, 0.7, 1.4, 2.0, 2.5, 3.0\}$}
We now turn our attention to the IPL-critical length and optimal-drift length for different appendage shapes. Under the constraint of elliptic shapes only, we investigate the aspect ratios $A = 0.2$, $0.7$, $1.4$, $2.0$, $2.5$ and $3.0$. Turn and drift angles for all aspect ratios except for $A = 0.7$ and $A = 3.0$ are shown in Fig.~\ref{fig:all-app-turnDriftVsLength}. For the aspect ratio $A = 3.0$ the IPL instability was not observed. The first column of frames in Fig.~\ref{fig:all-app-turnDriftVsLength} shows schematically the appendage shape with length $L_{\max}$, which produces the largest drift angle \revAll{between appendages with the given aspect ratio}. We observe that for small aspect ratios ($A = 0.2$, Fig.~\ref{fig:all-app-turnDriftVsLength}$a$) increasing appendage length $L$ means extending appendage mainly in the $x$-direction. However, for larger aspect ratios ($A = 1.4$, $2.0$ and $2.5$ in Figs.~\ref{fig:all-app-turnDriftVsLength}$d$, \ref{fig:all-app-turnDriftVsLength}$g$ and \ref{fig:all-app-turnDriftVsLength}$j$) increasing  length $L$ leads to appendage extension in the $y$-direction as well. Such a freedom is not present in 2D situation and it is likely to introduce flow structures in the wake, which do not exist in 2D. 

Comparing the turn angle curves in the second column (Figs.~\ref{fig:all-app-turnDriftVsLength}$b$, \ref{fig:all-app-turnDriftVsLength}$e$, \ref{fig:all-app-turnDriftVsLength}$h$, \ref{fig:all-app-turnDriftVsLength}$k$ and \ref{fig:single-app-turnDriftVsLength}$b$) one can observe that for intermediate aspect ratios ($A = 1.4$ and $A = 2.0$), the turn angle approaches some finite value at zero length, suggesting that very small appendages, which has some surface area to interact with flow, will generate a significant turn. In contrast, for small aspect ratios (Figs.~\ref{fig:all-app-turnDriftVsLength}$b$ and \ref{fig:single-app-turnDriftVsLength}$b$) the turn angle rapidly approaches zero, because the appendage geometry is approaching a needle-like shape (thickness starting to be comparable to width and length of the appendage), which most likely interacts with the back flow region in a different way compared to the planar appendage. \revAll{For example, it is possible that the needle-like appendage doesn't have large enough surface area to generate significant torque that would turn the appendage towards the wake attachment angle and therefore the turn angle for short appendages approaches zero.} On the other hand, for the largest aspect ratio  (Fig.~\ref{fig:all-app-turnDriftVsLength}$k$) the zero turn angle for short
appendage can be explained by the fact that the surface area of the appendage exposed to the back flow generated by the sphere is not sufficient for large enough destabilizing torque (that would lead to IPL instability) to develop. By looking at the drift angle curves in the third column (Figs.~\ref{fig:all-app-turnDriftVsLength}$c$, \ref{fig:all-app-turnDriftVsLength}$f$, \ref{fig:all-app-turnDriftVsLength}$i$, \ref{fig:all-app-turnDriftVsLength}$l$ and \ref{fig:single-app-turnDriftVsLength}$c$), one can observe that for all appendage shapes there exists an optimal length $L_{\max}$, which yields the largest drift angle. 
\begin{table}[b]
\revAll{
\begin{center}
\begin{tabular}{p{10mm} p{30mm} p{13mm} p{10mm} p{10mm} p{10mm} p{10mm} p{10mm}}
$A$ & $L_c$ & $L_{\max}$ & $\alpha_{\max}$ & $\theta_{\max}$ & $C_\scr{drag}$ & $C_\scr{drag,p}$ & $C_\scr{drag,\mu}$ \\
\hline
$0.2$ & $\left( 0.875 \pm 0.025 \right)\,D$ & $0.75\,D$ & $3.4^{\circ}$ & $12.1^{\circ}$ & $0.7482$ & $0.4113$ & $0.3369$ \\
$0.7$ & $\left( 0.825 \pm 0.025 \right)\,D$ & $0.70\,D$ & $4.8^{\circ}$ & $11.3^{\circ}$ & $0.7530$ & $0.4090$ & $0.3440$ \\
$1.4$ & $\left( 0.675 \pm 0.025 \right)\,D$ & $0.50\,D$ & $3.6^{\circ}$ & $11.9^{\circ}$ & $0.8189$ & $0.4195$ & $0.3994$\\
$2.0$ & $\left( 0.520 \pm 0.020 \right)\,D$ & $0.28\,D$ & $2.9^{\circ}$ & $12.4^{\circ}$ & $0.8532$ & $0.4390$ & $0.4142$ \\
$2.5$ & $\left( 0.380 \pm 0.020 \right)\,D$ & $0.16\,D$ & $2.2^{\circ}$ & $8.1^{\circ}$ & $0.8607$ & $0.4436$ & $0.4171$ \\
$3.0$ &  -- & -- & $0.0^{\circ}$ & -- & -- & -- & --
\end{tabular}
\end{center}
\caption{Critical length $L_c$ and maximum drift angle $\alpha_{\max}$ for all
aspect ratios $A$ of the elliptic appendage. $L_\scr{max}$ is appendage length,
which yields maximum drift angle $\alpha_{\max}$ at an equilibrium
turn angle $\theta_{\max}$. The corresponding drag
coefficient $C_\scr{drag}$ is also reported (for the configuration, which
gives the largest drift). In addition, both pressure drag coefficient $C_\scr{drag,p}$ and viscous
drag coefficient $C_\scr{drag,\mu}$ is reported. \label{tab:crit-L-optimal-drift}}
}
\end{table}

In order to summarize the results, we extract the critical length for the IPL instability $L_c$, the appendage length $L_{\max}$ yielding maximum drift angle, the largest drift angle $\alpha_{\max}$, the corresponding \revAll{turn angle $\theta_{\max}$ and the corresponding total} drag coefficient $C_{\scr{drag}}$ from simulations of all appendage shapes. \revAll{We also check the pressure drag $C_\scr{drag,p}$ and viscous drag $C_\scr{drag,\mu}$ contributions separately.} We present the findings in Tab.~\ref{tab:crit-L-optimal-drift}. From these results one can observe that there exists an upper limit on the aspect ratio for elliptic appendages anchored to the center of sphere somewhere between $A = 2.5$ and $A = 3.0$.  The lower bound, however, we were not able to identify in this work. Looking at the trend of $L_c$ between different aspect ratios, one can state that smaller aspect ratio leads to larger critical length. The appendage with a small aspect ratio is localized at the center of the wake, where the back flow region is the longest. Therefore the appendage can extend longer compared to other appendages with higher aspect ratios without being exposed to forward flow.

\begin{figure}
  \begin{center}
  \includegraphics[width=0.7\linewidth]{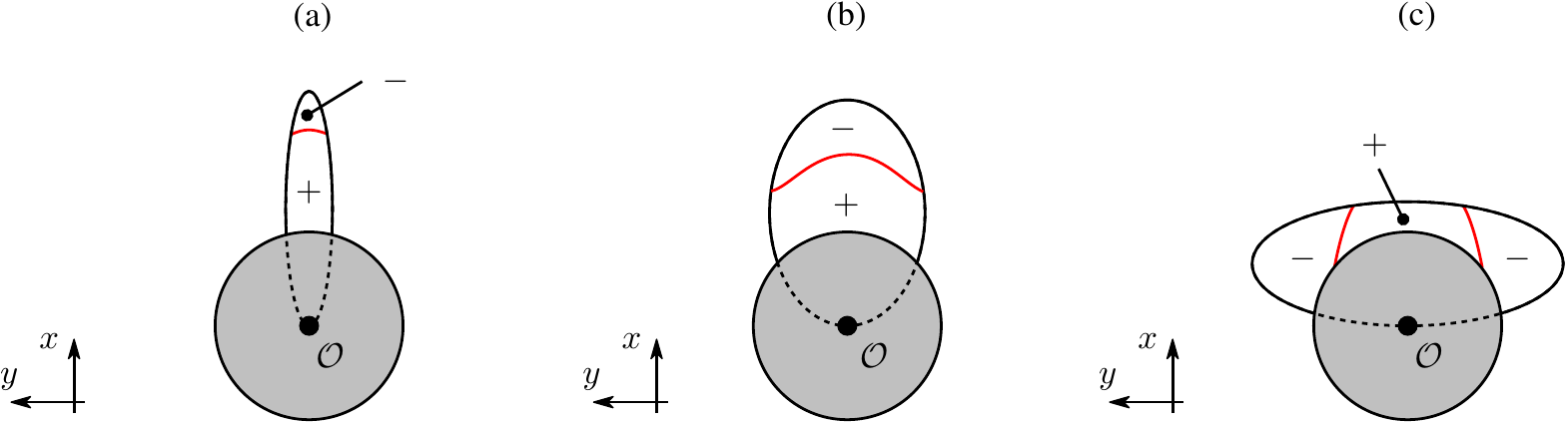} 
  \end{center}
  \caption{
This figure shows authors' impression of regions of positive ``$+$'' and
negative ``$-$'' normal force of the elliptic appendage. Frame (a) shows an appendage with small aspect ratio $A = 0.2$, for which the appendage is extending out of the
back flow region in $x$ direction. Frame (b) shows an appendage with
moderate aspect ratio $A = 0.7$, which also extends out of the back flow region
in $x$ direction. Right frame (c) shows an appendage with large aspect ratio $A = 2.5$, which extends out of the back flow region in $y$ direction. Out of all these shapes, the one with $A = 0.7$ shows the largest surface area (region ``$+$'') exposed to destabilizing force. \label{fig:illustrate-plusMinus-smallLargeA}}
\end{figure}

As for the drift angle, one can observe that there exists an optimal aspect ratio, which yields the largest drift angle compared to other appendage shapes investigated. The optimal aspect ratio found in this work is $A_{\textrm{opt}} = 0.7$; with a finer sampling of aspect ratios, the optimal aspect ratio should be somewhere in the interval $A_{\textrm{opt}} \in \left(0.4, 1.1 \right)$. The reason that this particular and small aspect ratio induces the largest drift can be understood as follows. For smaller aspect ratios, say $A = 0.2$,  the appendage is relatively thin and long, that is, the surface area is small.  We illustrate our impression of the positive and negative force regions for this appendage in Fig.~\ref{fig:illustrate-plusMinus-smallLargeA}$a$. Consequently only a small portion of the energy from the recirculation region can be used. Now, on the other hand, for larger aspect ratios, such as $A = 2.5$, the appendage is rapidly extending from the back flow region in the $y$ direction and by doing so, it produces large stabilizing forces. Also, for this configuration, the surface area of the appendage that is exposed to the back flow is relatively small, as sketched in Fig.~\ref{fig:illustrate-plusMinus-smallLargeA}$c$. However, the shape of the appendage with aspect ratio $A = 0.7$ is such that, while the appendage is extending out of the back flow region in the $x$ direction similarly as $A = 0.2$, the appendage is wide enough in $y$ direction to take advantage of the width of the back flow region and recover larger part of the energy present in the recirculation bubble. 
\revAll{
It is interesting to point out, that the turn angles $\theta_{\max}$ (see
Tab.~\ref{tab:crit-L-optimal-drift}), corresponding to the maximum drift angle, are very similar
over wide range of aspect ratios $A \in \left( 0.2, 2.0 \right)$. The reason for this similarity
we leave as an open question.
}

\begin{figure}
  \begin{center}
  \includegraphics[width=0.7\linewidth]{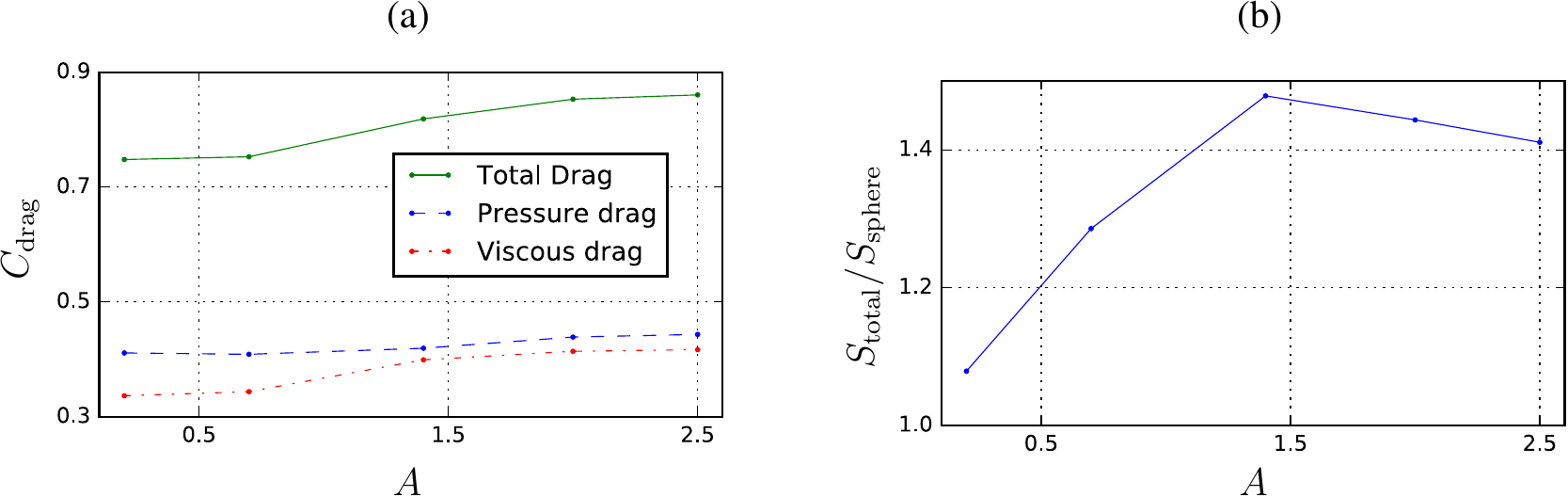} 
  \end{center}
  \caption{\revAll{
  In frame (a) we show the total drag $C_\scr{drag}$ around sphere with an appendage
  at an equilibrium state division in pressure drag $C_\scr{drag,p}$
  and viscous drag $C_\scr{drag,\mu}$ for all considered aspect ratios $A$. 
  In frame (b) we show the total surface area of the body $S_\scr{total}$ relative
  to the surface area of sphere alone $S_\scr{sphere}$. Value of
  $S_\scr{total}/S_\scr{sphere} = 1.0$ would show that the body has a total surface
  area, which is the same as for the sphere.} \label{fig:drag-pressure-viscous}}
\end{figure}

Finally, we can see from Tab.~\ref{tab:crit-L-optimal-drift}  that increasing the aspect ratio leads also to an increased drag coefficient. For example, if one changes aspect ratio from $A_{\scr{opt}} = 0.7$ to $A = 2.0$, the drift angle is reduced by $40\%$ but at the same time the drag is increased by $13\%$.
\revAll{
This increase of the drag can be partially attributed to the increase of the
total surface area of the
body due to the addition of the appendage and corresponding increase in viscous drag.
In Fig.~\ref{fig:drag-pressure-viscous}$b$ we
show how the ratio between the total surface area and that of
sphere alone $S_\scr{total}/S_\scr{sphere}$ varies between
different aspect ratios.
%In the
%computation of the surface area, we have neglected the small contribution from the sides
%of the appendage with the thickness of $0.02\,D$.
The total surface area of the body has been obtained by adding twice the surface area
of the extruding part of the ellipse to the surface area of the sphere.
From Fig.~\ref{fig:drag-pressure-viscous}$b$
one can note that the total surface area of the body
is increased by around $40\%$ when increasing the aspect ratio from $0.2$ to $1.4$.
The pressure and viscous
drag contributions are presented in
last columns of Tab.~\ref{tab:crit-L-optimal-drift} and
plotted in Fig.~\ref{fig:drag-pressure-viscous}$a$. There we see that the viscous drag
increases more rapidly compared to pressure drag
when going from aspect ratio $0.2$ to $1.4$.
This rapid increase corresponds to aspect ratio interval,
in which the total surface area of the body
increases the most. Therefore the bulk of the drag increase can indeed be explained by
the increased importance of the viscous drag.
%This increase of the drag
Larger values of the total drag for higher aspect ratios than $1.4$
most likely could be attributed to increase of the projected area of the body,
or in other words, to the extension of the appendage
outside of the sphere wake (see Fig.~\ref{fig:all-app-turnDriftVsLength}$g$).}
An increased drag could be advantageous, for example, in seed dispersion, where an increased drag increases time spent in air and maximizes the effect that surrounding wind can have on the free-fall trajectory.

\section{Conclusions} \label{sec:discuss}
In this work, we have used numerical simulations and the understanding of the physical mechanisms obtained from the simple model presented by L\={a}cis et al.\cite{Lacis2014driftSymmetry} to characterize the forces on three-dimensional bodies, consisting of a sphere and a planar appendage in the shape of an ellipse. Our main results can be summarized in two points. First, the IPL instability exists for a wide range of appendages, and the physical mechanism in three-dimensional setting is exactly the same as in two-dimensional setting, despite the fact that the wake structure is much more complex. We have explained physically the behaviour of forces for different turn angles by looking at the normal forces on the appendage, the resulting torque from those forces and the projection of those forces in the drift direction. Second, through a systematic parametric investigation of aspect ratios and appendage lengths, we were able to identify the appendage ($A=0.7, L=0.7)$, which leads to the largest drift angle $(\alpha=4.8^\circ)$ of a body that would be free to fall \revAll{at Reynolds number $Re = 200$}.  We have  explained this optimum solution by comparing different surface areas exposed to back flow region. We concluded that for small aspect ratios the surface area exposed to back flow region is small, because the shape of the appendage is thin. For large aspect ratios, on the other hand, the surface area exposed to back flow region is small, because the extension of the appendage in $y$ direction reduces the effect of IPL instability very rapidly.

%\revAll{
%Note that the currently investigated Reynolds number $Re = 200$ is very close to
%the critical Reynolds number of symmetry breaking instability in the
%wake behind the sphere alone, which is either $Re^{SS} = 212$ for
%fixed sphere or $Re^{OS} = 206$ for freely rotating sphere \cite{fabre2016flow}.
%This gives rise to a question, whether there exists some sort of interaction
%or competition between the IPL instability and
%the symmetry breaking instability in the wake of the sphere.
%In order to answer this question, the future work should provide the investigation
%of IPL instability for various elliptic appendages
%for different values of Reynolds numbers, both further away from the symmetry
%breaking of the wake behind the sphere and within the values of that symmetry breaking.
%}

Further investigations of how the energy loss in formation of the recirculation behind a sphere or other bluff bodies can be beneficially exploited to enable innovative passive control techniques. Already, the very simple configuration studied in this paper shows that the induced side force by the appendage is significant, and can be considered as a means to passively control the dynamics of bluff body wakes. For example, one may add appendages to spherical particles to increase the dispersion, although further studies including particle-particle interaction remains to be conducted.  We also observed that larger aspect ratios yielded larger drag coefficients, which can be beneficial to spend longer time in air during free fall motion, while still exhibiting some IPL drift.  In future studies, we will consider more complex elastic appendages such as appendages with holes and multiple planes similar to Fallopia seeds\cite{rouifed2011achene} as well as more complex \revAll{steady or} unsteady wakes.
\revAll{For example, the wake behind sphere alone already at $Re^{SS} = 212$
for fixed sphere or at $Re^{OS} = 206$ for freely rotating sphere\cite{fabre2016flow}
exhibits symmetry breaking. It would be interesting to investigate
in detail the interaction between this symmetry breaking and IPL instability.}
The direct numerical simulation of such complex bodies falling/rising freely is very challenging, and it is likely to be -- at least initially -- modeled by a more sophisticated model than the one presented in~\cite{Lacis2014driftSymmetry}. One possible direction is to extend the model of L\={a}cis et al.\cite{Lacis2014driftSymmetry} from 2D to 3D, while another possibility is to investigate the weakly non-linear model by Tchoufag et al.\cite{tchoufag2015weakly}. The latter model is able to predict oblique falling paths of disks and bubbles and could be extended to more complex bodies in order to exploit the IPL instability.

\begin{acknowledgments}
U.L. and S.B. acknowledges the financial support from the Swedish Research Council (VR-2014-5680) and the G{\"o}ran Gustafsson foundation. S.O. and A.M. thanks the financial support for the computational infrastructure from the RITMARE project and the PRIN 2012 project (no. D38C13000610001) funded by the Italian Ministry of Education.
\end{acknowledgments}

\bibliography{/home/ugis/Documents/references_UgisL.bib}
%\bibliography{./references_UgisL}

\end{document}